\documentclass[10pt,twocolumn]{article}

\usepackage[utf8]{inputenc}
\usepackage{amsmath}
\usepackage{graphicx}
\usepackage[authoryear,sort]{natbib}

\RequirePackage{xcolor}
\definecolor{goldenbrown}{rgb}{0.6, 0.4, 0.08}
\RequirePackage[colorlinks,linkcolor=goldenbrown,citecolor=goldenbrown,urlcolor=goldenbrown]{hyperref}

\usepackage{fancyhdr}
\RequirePackage{amsmath,amsthm,amsfonts,amssymb}
\usepackage{caption}
\usepackage[margin=.5in,top=0.7in,bottom=.7in]{geometry}
\usepackage{algorithm}

\usepackage[explicit]{titlesec} 
\titleformat{\section}{\normalfont\large\bf}{\thesection}{1em}{#1.}
\titleformat{\subsection}[runin]
  {\normalfont\normalsize\bf}{\thesubsection}{1em}{#1.}
\titleformat{\subsubsection}[runin]
  {\normalfont\normalsize\it}{\thesubsubsection}{1em}{#1.}

\date{}

\newcommand{\iid}{\stackrel{\rm i.i.d.}{\sim}}
\newcommand{\sumn}{\sum\limits_{i=1}^n}

\usepackage{float} 

\newlength\FHoffset
\fancyheadoffset{\FHoffset}
\title{\vspace{-10mm}{\bf \large PROBABILISTIC GRAPHICAL MODELS IN ASTRONOMY}}
\author{\sc Abigail Sheerin and  Giuseppe Vinci\footnote{Corresponding author; E-mail: gvinci@nd.edu}\\{\footnotesize Department of Applied and Computational Mathematics and Statistics}\vspace{-2mm}\\  {\footnotesize  University of Notre Dame, Notre Dame, Indiana, USA}}

\begin{document}

\maketitle

\paragraph{\bf Abstract.} The field of astronomy is experiencing a data explosion driven by significant advances in observational instrumentation, and classical methods often fall short of addressing the complexity of modern astronomical datasets. Probabilistic graphical models offer powerful tools for uncovering the dependence structures and data-generating processes underlying a wide array of cosmic variables. By representing variables as nodes in a network, these models allow for the visualization and analysis of the intricate relationships that underpin theories of hierarchical structure formation within the universe. We highlight the value that graphical models bring to astronomical research by demonstrating their practical application to the study of exoplanets and host stars.
\vspace{2mm}

\noindent {\small {\sc keywords}: astrostatistics, exoplanets, Gaussian graphical model, multiple hypothesis testing, regularization, star.}

\section{Introduction}\label{introduction}
Astrophysical theory has long been central to understanding the cosmos, from planetary formation to the evolution of the universe. However, astrophysics requires statistical models for the proper interpretation of astronomical data to test cosmological theories \citep{feigelson2021twenty}. The study of our universe through statistical analysis, \textit{astrostatistics},  relies on analyzing diverse types of data about  galaxies, stars, planets, and other celestial objects. Key among these data types are photometry \citep{eyer2019multivariate}, spectrophotometry \citep{christian1977multivariate},  spectroscopy \citep{lasue2012remote}, galactic properties \citep{corbin2001multivariate}, luminosity \citep{fabbiano2002multivariate}, age and metallicity \citep{carraro1998galactic}, rotation and velocity \citep{das2015multivariate}, element abundance \citep{nissen2018high}, and geological mapping \citep{bhatt2019global}. 

The field of astrostatistics relies on diverse statistical and machine learning methods for astrophysical data analysis. These methods can be broadly categorized into unsupervised and supervised learning. Unsupervised learning techniques allow for the discovery of patterns or structures in data without prespecified labels. Various unsupervised learning methods for dimensionality reduction, clustering, and data completion have been applied in astrostatistics. These methods include: principal component analysis \citep{efstathiou1984multivariate, fiorentin2007estimation, bailey2012principal,  eyer2019multivariate}; independent component analysis \citep{Maino2001,Lu2006,Allen2013,das2015multivariate,Akutsu2020,Pati2021}; sparse dictionary learning \citep{vinci2014estimating,Constantina2017}; and k-means clustering \citep{Chattopadhyay2010,das2015multivariate,fraix2015multivariate,Turner2018,Nascimento2022,ikotun2023k}. 

Supervised learning methods, including regression and classification, are  widely used in astrostatistics for prediction. While they help extract meaningful patterns from complex datasets like unsupervised methods, supervised learning methods require a response variable and predictors. These methods provide powerful tools for analyzing astronomical data, enabling discoveries in areas such as galaxy classification, stellar dynamics, and elemental mapping \citep{feigelson2012modern}. 
Various supervised learning methods have been applied in astrostatistics. These methods include: linear regression \citep{Feigelson1992,Magorrian1998,Kelly2007,Riess2016,bhatt2019global}; generalized linear models \citep{AndreonHurn2010,deSouza2015Binomial,deSouza2015NegBin,Elliott2015Gamma}; and random forests, k-Nearest Neighbors, support vector machines, and deep learning  for automated classification of galaxy morphologies and stars \citep{freeman2013new,dieleman2015rotation,kim2017stargalaxy,baumstark2024spiral,nemade2024morphological}.

The field of astronomy has been experiencing a data explosion driven by significant advancements in observational instruments. These new tools allow for more detailed and frequent observations, resulting in an unprecedented and exponentially growing volume of data, with sample sizes ranging from dozens to billions of objects \citep{cisewski2019special, feigelson2021twenty}. This data surge has propelled astronomy into the ``big data'' era \citep{zhang2015astronomy}, exceeding the capabilities of traditional statistical frameworks. This massive influx of information presents considerable challenges across the entire data lifecycle. From cleaning and integration to processing, indexing, mining, visualization, and analysis, each stage is impacted. Consequently, the statistical methods traditionally applied in astronomy require continuous improvements to effectively handle the scale and complexity of this new data landscape \citep{feigelson2012modern}.

Probabilistic graphical models \citep{lauritzen1996graphical} are powerful tools for describing the dependencies among many random variables. In a graphical model, the dependence structure of $d$ random variables $X_1,...,X_d$ is encoded by a graph $G=(V,E)$, where the vertices or nodes $V=\{1,...,d\}$ represent the variables, and the edges in $E$ connect nodes to reflect conditional dependence relationships.  Given the data, we are interested in estimating $E$. 

Graphical models have been applied in numerous fields, including neuroscience \citep{ortiz2015exploratory,vinci2018adjusted,vinci2018adjustedB,zhu2018sparse}, genomics \citep{dobra2004sparse,yin2011sparse,chun2013joint,gan2022correlation}, proteomics \citep{wang2016fastggm}, metabolomics \citep{krumsiek2011gaussian}, forensic science \citep{dawid2007object,xu2024forensic}, and finance \citep{carvalho2007dynamic}. Despite their potential, graphical models remain underutilized in astrostatistics.

In this article, we illustrate the application of probabilistic graphical models in astronomy. 
In Section~\ref{sec:gms}, we provide an overview of probabilistic graphical models and methods for their estimation.  In Section~\ref{sec:data}, we apply graphical models to detect dependencies among variables concerning exoplanets and host stars. Finally, in Section~\ref{sec:disc}, we discuss our results and future research directions.

\section{Probabilistic Graphical Models}\label{sec:gms}
Probabilistic graphical models are multivariate probability distributions of random variables $X_1,...,X_d$ with a dependence structure represented by a {\it graph} $G=(V,E)$, where $V=\{1,...,d\}$ is a set of {\it nodes} (or vertices) representing $d$ random variables, and $E\subset V\times V$ is a set of {\it edges} encoding dependence relationships among the variables. Two important classes of graphical models are  \textit{undirected} and \textit{directed} graphical models.

\begin{figure}[ht]
    \centering    
    \includegraphics[width=1\columnwidth]{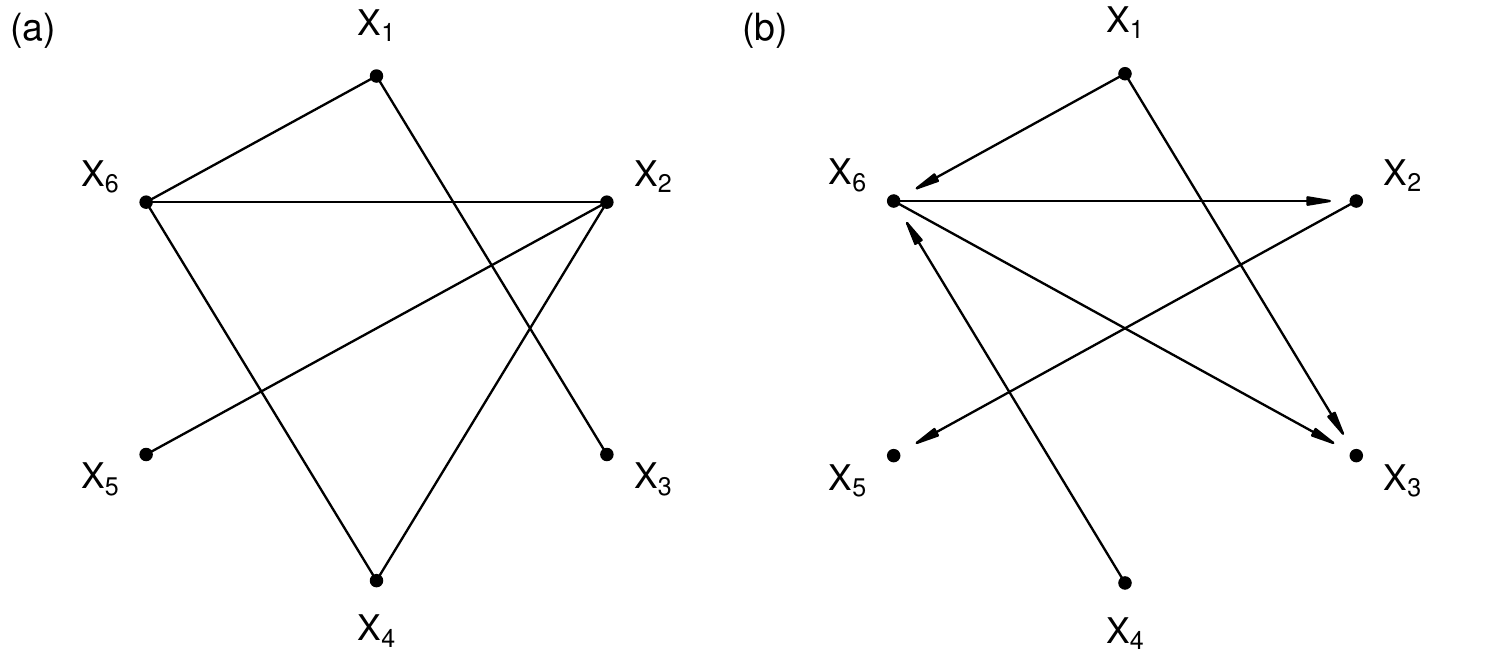}
\caption{(a): An undirected graph where any two variables are not connected by an edge if and only if they are independent conditionally on all other variables. For example, $X_1$ and $X_5$ are independent conditionally on all other variables. (b): A directed acyclic graph (DAG) where a variable $X_i$ is independent of a variable $X_j$ conditionally on the parents of $X_i$ if and only if $X_j$ is neither a parent nor a descendant of $X_i$. For example, $X_6$ is the only parent of $X_2$ so, conditionally on $X_6$, $X_2$ is independent of $(X_1,X_3,X_4)$, but it is dependent on its descendant $X_5$.}
    \label{fig:graphtypes}
\end{figure}

\subsection{Undirected Graphs}
A graph $G=(V,E)$ is {\it undirected} if the edge set $E$ is symmetric, i.e.,  $(i,j)\in E \Leftrightarrow (j,i)\in E$ \citep{lauritzen1996graphical}. Figure~\ref{fig:graphtypes}(a) shows an example of an undirected graph. We say that a $d$-variate random vector $X=(X_1,...,X_d)^T$ forms a Markov Random Field (MRF) with respect to an undirected graph $G=(V,E)$ if the \textit{pairwise Markov property} holds: $X_i$ and $X_j$ are independent conditionally on all other variables if and only if $(i,j)\notin E$. If the distribution of $X$ is positive, then the \textit{local} and  \textit{global} Markov properties are also satisfied: any variable $X_i$ is conditionally  independent of all other variables given  $X_{\mathcal{N}_E(i)}=(X_k)_{k\in\mathcal{N}_E(i)}$, where $\mathcal{N}_E(i)=\{j:(i,j)\in E\}$ are the neighbors of $i$; and any two distinct subsets of variables are independent conditionally on a separating set, i.e., the set of variables that, if removed from the graph, would disconnect the two variable sets. For example, in Figure~\ref{fig:graphtypes}(a), $X_1$ and $X_2$ are independent conditionally on all other variables (pairwise Markov property); $X_1$ is independent of $(X_2,X_4,X_5)$ conditionally on its neighbors $\mathcal(X_3,X_6)$ (local Markov property); and $(X_2,X_5)$ are independent of $(X_1,X_3)$ conditionally on $(X_4,X_6)$ (global Markov property). 

A well established MRF is the Gaussian Graphical Model (GGM), where $X\sim N(\mu,\Theta^{-1})$, $\mu\in\mathbb{R}^d$ is the mean vector, $\Theta\succ 0$ is a $d\times d$ symmetric positive definite \textit{precision matrix}, and the edge set is $E=\{(i,j)\in V\times V: i\neq j, \Theta_{ij}\neq 0\}$. Equivalently, the edge set in a GGM is
\begin{equation}
E~=~\left\{(i,j)\in V\times V: i\neq j, ~\rho_{ij}\neq 0\right\}
\end{equation}
where 
\begin{equation}\label{eq:parcor}
\rho_{ij} ~=~ -\Theta_{ij}\left(\Theta_{ii}\Theta_{jj}\right)^{-1/2}
\end{equation}
is the \textit{partial correlation} of $(X_i,X_j)$, i.e., the correlation between $(X_i,X_j)$ conditionally on all the other $d-2$ variables. Note that $-1<\rho_{ij}<1$, just like the \textit{marginal Pearson's correlation}
\begin{equation}
    r_{ij}~=~\Sigma_{ij}(\Sigma_{ii}\Sigma_{jj})^{-1/2},
\end{equation}
where $\Sigma=\Theta^{-1}$ is the \textit{covariance matrix} of $X$. However, $r_{ij}$ neglects the effects of other variables on the dependence between $(X_i,X_j)$, so it is possible to have $r_{ij}\neq 0$ while $\rho_{ij}=0$, or even $r_{ij}<0$ and $\rho_{ij}>0$, and vice versa. 

In real data analysis, the primary goal is to use data to estimate $G=(V,E)$, which, in a GGM, reduces to estimating the support of $\Theta$ or, equivalently, identifying all pairs $(i,j)$ such that $\rho_{ij}\neq 0$.

\subsubsection{Estimating a GGM via multiple hypothesis testing}
Suppose we observe $n$ independent and identically distributed (i.i.d.) data vectors $X^{(1)},...,X^{(n)}\iid N(\mu,\Theta^{-1})$, where $X^{(r)}=(X^{(r)}_1,...,X^{(r)}_d)^T$. An approach to estimating $E$ requires first obtaining the \textit{maximum likelihood estimate} (MLE) of $\left(\mu,\Theta\right)$,
\begin{equation}\label{eq:mle}
(\hat\mu,\hat\Theta) ~=~ \underset{\mu\in\mathbb{R}^d,\Theta\succ 0}{\arg\max}~ L(\mu,\Theta;X^{(1)},...,X^{(n)}),
\end{equation}
where 
\begin{equation}
L(\mu,\Theta;X^{(1)},...,X^{(n)})~=~\prod_{i=1}^n p(X^{(i)};\mu,\Theta)
\end{equation} 
is the likelihood function, and 
\begin{equation}
p(x;\mu,\Theta)~=~\sqrt{\frac{\det\Theta}{(2\pi)^d}} \exp\left(-\tfrac{1}{2} (x-\mu)^T\Theta (x-\mu)\right)
\end{equation}
is the probability density function (p.d.f.) of $N(\mu,\Theta^{-1})$. If $n>d$, we have $\hat\mu=\frac{1}{n}\sumn X^{(i)}$ and $\hat\Theta=\hat \Sigma^{-1}$, where 
\begin{equation}
    \hat \Sigma ~=~ \frac{1}{n}\sumn (X^{(i)}-\hat\mu)(X^{(i)}-\hat\mu)^T
\end{equation}
is the $d\times d$ sample covariance matrix. 

The matrix $\hat\Theta$ is dense almost surely, so it does not readily provide us with an estimate of the graph structure of $\Theta$. Thus, we use hypothesis testing to assess whether $\rho_{ij}\neq 0$ based on the partial correlation MLE (or sample partial correlation)
\begin{equation}\label{eq:parcorhat}
\hat\rho_{ij}~=~-\hat\Theta_{ij}(\hat\Theta_{ii}\hat\Theta_{jj})^{-1/2},
\end{equation}
for $1\le i<j\le j$. For large $n$, we have 
\begin{equation}\label{eq:fisherasymp}
g\left(\hat\rho_{ij}\right)~\approx~ N\left(g(\rho_{ij}),~s^2_n
\right),
\end{equation}
where $s^2_n=(n-d-1)^{-1}$, and $g(\omega)=\frac{1}{2}\log\left((1+\omega)/(1-\omega)\right)$ is the Fisher transformation, a monotone increasing function that maps the interval $(-1,1)$ onto the real line $\mathbb{R}=(-\infty,\infty)$, which is the support of the Gaussian distribution \citep{fisher1924distribution}. Thus, to test the hypothesis $H_{0,ij}:\rho_{ij}=0$ versus $H_{1,ij}:\rho_{ij}\neq 0$, we compute the p-value
\begin{equation}
P_{ij} ~=~ 2\left(1-\Phi\left(|g(\hat\rho_{ij})|s^{-1}_n\right)\right),
\end{equation}
where 
\begin{equation}\label{eq:phiz}
\Phi(z) ~=~ \int_{-\infty}^z(2\pi)^{-1/2}e^{-\omega^2/2}d\omega
\end{equation}
is the cumulative distribution function (c.d.f.) of the standard normal (Gaussian) distribution $N(0,1)$. Finally, we apply methods for multiple hypothesis testing to obtain estimates of the edge set $E$. 

Two possible approaches for multiple hypothesis testing are family-wise type I error rate control and false discovery rate control. 
The \textit{family-wise type I error rate} (FWER) is the probability of (erroneously) discovering at least one edge while assuming that the true graph contains no edges. 
By applying the Bonferroni correction \citep{bonferroni1936teoria,dunn1961multiple}, we can define the edge set estimator
\begin{equation}\label{eq:bonferroniedgeset}
\hat E(\alpha) ~=~ \left\{(i,j)\in V\times V: i\neq j, P_{ij}\le\alpha/m\right\}, 
\end{equation}
where $\alpha\in (0,1)$ is the desired FWER level, and $m=\binom{d}{2}$ is the number of hypotheses. 
Equivalently, based on Equation~\eqref{eq:fisherasymp}, an approximate $1-\alpha$ confidence interval for $\rho_{ij}$ is
\begin{equation}\label{eq:fisherconfint} 
    C_{ij}(\alpha)~=~\left[g^{-1}\hspace{-1mm}\left(g(\hat\rho_{ij})-z_{\frac{\alpha}{2}} s_n\right),g^{-1}\hspace{-1mm}\left(g(\hat\rho_{ij})+z_{\frac{\alpha}{2}} s_n\right)\right],
\end{equation}
where $g^{-1}(z) = (\exp(2z)-1)/(\exp(2z)+1)$ is the inverse Fisher transformation, $z_{\alpha/2}=\Phi^{-1}(1-\alpha/2)$ is the $1-\alpha/2$ quantile of $N(0,1)$, and $\Phi^{-1}()$ is the inverse of the c.d.f. of $N(0,1)$ in Equation~\eqref{eq:phiz}. Thus, the edge set in Equation~\eqref{eq:bonferroniedgeset} can be rewritten as
\begin{equation}\label{eq:bonferroniedgeset2}
\hat E(\alpha) ~=~ \left\{(i,j)\in V\times V: i\neq j, 0\notin C_{ij}\big(\alpha/m\big)\right\} 
\end{equation}
The \textit{false discovery rate} (FDR) is the expected proportion of incorrect edges among all discovered edges. By applying the FDR control of \cite{benjamini2001control} for dependent tests, we can define the edge set estimator
\begin{equation}\label{eq:fdredgeset}
\hat E(\beta) ~=~ \left\{(i,j)\in V\times V: ~i\neq j, P_{ij}\le P^*_\beta\right\}, 
\end{equation}
where $P^*_\beta=P^{(k^*)}$ is the $k^*$-th smallest p-value among the $m=\binom{d}{2}$ p-values $\{P_{ij}\}_{1\le i<j\le d}$, 
\begin{equation}
    k^* ~=~\max\left\{k: P^{(k)}\le 
\beta k\left(m\sum\limits_{j=1}^m j^{-1}\right)^{-1}\right\},
\end{equation}
and $\beta\in (0,1)$ is the desired FDR level.

In the data analysis in Section~\ref{sec:data}, we will also estimate \textit{marginal correlation graphs} based on MLE and multiple hypothesis testing, i.e., graphs analogous to Equations~\eqref{eq:bonferroniedgeset} and \eqref{eq:fdredgeset} but based on the marginal correlation MLEs (or sample marginal correlations),
\begin{equation}\label{eq:corrmle}
\hat r_{ij}~=~ \hat\Sigma_{ij}(\hat\Sigma_{ii}\hat\Sigma_{jj})^{-1/2},
\end{equation}
rather than the partial correlation estimates $\hat\rho_{ij}$, and with $s^2_n=(n-3)^{-1}$. Since the marginal correlation $r_{ij}$ neglects the effects of other variables on the dependence between $(X_i,X_j)$, a correlation graph is typically much denser (if not fully connected) than a conditional dependence graph.

\subsubsection{Estimating a GGM via penalized MLE}
If the sample size $n$ is not large enough compared with the number of nodes $d$, the MLE $\hat\Theta$ may have undesirable statistical properties or may not even have a unique solution (for example, if $n<d$, $\hat\Sigma$ is not invertible). To circumvent the issues of the MLE, several {\it penalized MLE} methodologies have been proposed. A notable example is the \textit{Graphical LASSO} (GLASSO) \citep{yuan2007model}, which is the solution of
\begin{equation}\label{eq:glasso}
\hat\Theta(\lambda)~=~\underset{\Theta\succ 0}{\arg\max} ~~\ell_n(\Theta)-\lambda \Vert\Theta\Vert_{1,\rm off},
\end{equation}
where 
\begin{equation}
\ell_n(\Theta) ~=~ \ln L(\hat\mu,\Theta;X^{(1)},...,X^{(n)})
\end{equation}
is the log likelihood function maximized with respect to $\mu$ (maximizer $\hat\mu=\frac{1}{n}\sumn X^{(i)}$), and $\lambda$ is a tuning parameter controlling the sparsity effect of the $L_1$ norm 
$\Vert\Theta\Vert_{1,\rm off}:=\sum_{i\neq j}|\Theta_{ij}|$ on $\hat\Theta(\lambda)$. For a given value of $\lambda$, the GLASSO estimate of $E$ is
\begin{equation}\label{eq:glassoE}
\hat E(\lambda) ~=~ \left\{(i,j)\in V\times V: i\neq j, \hat\Theta_{ij}(\lambda)\neq 0\right\},
\end{equation}
which is generally sparser for larger values of $\lambda>0$. Note that, for $n>d$,  $\hat\Theta(0)=\hat\Theta$, the MLE in Equation~\ref{eq:mle}.

Thus, in GLASSO, graphical model selection reduces to selecting the tuning parameter $\lambda$. A well established approach to select $\lambda$ is the Extended Bayesian Information Criterion (EBIC; \cite{foygel2010extended}), which picks $\lambda$ by minimizing the risk function
\begin{equation}\label{eq:ebic}
    {\rm EBIC}(\lambda) ~=~ -2\ell_n(\tilde\Theta(\lambda))+\kappa(\lambda)\left(\log n+4\gamma\log d\right),
\end{equation}
where 
\begin{equation}
    \tilde\Theta(\lambda) ~=~ \underset{\Theta\succ 0,\Theta_{\hat E(\lambda)^c}=0}{\arg\max}~ \ell_n(\Theta)
\end{equation}
is the MLE of $\Theta$ constrained to have no edges beyond the GLASSO edge set $\hat E(\lambda)$ (Equation~\eqref{eq:glassoE}), $\kappa(\lambda)=|\hat E(\lambda)|/2$ is the number of edges in $\hat E(\lambda)$, and $\gamma$ is a parameter that we set equal to $0.5$ as recommended by the authors.

GLASSO is guaranteed to have a unique solution if $\lambda>0$ \citep{ravikumar2011high}, and various algorithms have been developed to estimate graphs with thousands of nodes efficiently \citep{yuan2007model,friedman2008sparse,hsieh2011sparse,mazumder2012graphical,hsieh2013big}. Furthermore, the excellent statistical performance of GLASSO has been demonstrated under numerous settings \citep{rothman2008sparse,yuan2010high,ravikumar2011high}. 

\subsubsection{Estimating a GGM from non-Gaussian data} Multivariate data often do not follow a Gaussian distribution, and other graphical models may be more appropriate. However, GGM estimation is computationally convenient, and various approaches exist to align empirical data with GGM assumptions. Some approaches include applying suitable monotonic transformations, such as the logarithm or the square root \citep{vinci2018adjustedB}, or embedding the GGM within a hierarchical model \citep{vinci2018adjusted}. Alternatively, we can transform the data \textit{nonparametrically} \citep{liu2012high} as follows. First, we define the empirical c.d.f.~of $X_i$
\begin{equation}
\hat F_i(x) ~=~ \frac{1}{n}\sum_{r=1}^n I(X_i^{(r)}\le x),
\end{equation}
where $I()$ is the indicator function, and its Winsorized version 
\begin{equation}
 \tilde F_i(x)~=~ \left\{\begin{array}{cc}
      \delta_n, & \text{if } \hat F_i(x)<\delta_n  \\
      1-\delta_n, & \text{if } \hat F_i(x)>1-\delta_n  \\
     \hat F_i(x), & \text{otherwise}
 \end{array}\right.
\end{equation}
where $\delta_n=(n+1)^{-1}$. 
Then, we compute the \textit{normal scores} $\tilde X_i^{(1)},...,\tilde X_i^{(n)}$, where
\begin{equation}\label{eq:normalscore}
    \tilde X_i^{(r)} ~=~ \Phi^{-1}\left(\tilde F_i(X_i^{(r)})\right)
\end{equation}
and $\Phi^{-1}(\cdot)$ is the inverse of the c.d.f.~of the standard normal (Gaussian) distribution $N(0,1)$ in Equation~\eqref{eq:phiz}. This transformation makes $ \tilde X_i^{(r)}$ approximately follow $N(0,1)$. 
Finally, we estimate a GGM based on the normal scores of the data. We apply this nonparametric approach in our data analysis in Section~\ref{sec:data}. 

\subsection{Directed Graphs}
A graph $G=(V,E)$ is {\it directed} if the edge set $E$ consists of ordered pairs of nodes, where $(i,j)\in E$ indicates $i\rightarrow j$, i.e., there is an {\it edge-arrow} pointing from the {\it parent} node $i$ to the {\it child} node $j$ \citep{lauritzen1996graphical}. Figure~\ref{fig:graphtypes}(b) shows an example of a directed graph. 
A \textit{directed path} is a set of arrows all pointing in the same direction and linking an \textit{ancestor} node $i$ to a \textit{descendant} node $j$, for example, $i\to k\to s \to j$. We say that the triple $(i,j,k)$ forms a \textit{collider} at $j$ if $i\rightarrow j \leftarrow k$; this collider is \textit{unshielded} if  $i$ and $k$ are not connected by any edge; otherwise, it is \textit{shielded}. For example, in Figure~\ref{fig:graphtypes}(b), the triple $(1,6,4)$ forms an unshielded collider at $6$, while the triple $(1,3,6)$ forms a shielded collider at $3$. 

\subsubsection{DAGs}
A \textit{directed acyclic graph} (DAG) is a directed graph that contains no cycles, i.e., directed paths that start and end with the same node. We say that a DAG $G=(V,E)$ represents the probability distribution of a random vector $X=(X_1,..,X_d)^T$ if the joint probability density function (p.d.f.)~or probability mass function (p.m.f.) of $X$ can be expressed as 
\begin{equation}
p_X(x_1,...,x_d)~=~\prod_{i=1}^d p_{X_i\mid X_{\pi(i)}=x_{\pi(i)}}(x_i),
\end{equation}
where 
\begin{equation}
p_{X_i\mid X_{\pi(i)}=x_{\pi(i)}}(x_i)~=~ \frac{p_{X_i,X_{\pi(i)}}(x_i,x_{\pi(i)})}{p_{X_{\pi(i)}}(x_{\pi(i)})}
\end{equation}
is the conditional p.d.f.~or p.m.f.~of $X_i$ given $X_{\pi(i)}=x_{\pi(i)}$, and $\pi(i)=\{j:(j,i)\in E\}$ is the set of parents of node $i$. It can be shown that $X_i$ is independent of $X_j$ conditionally on $X_{\pi(i)}$ if and only if $j$ is neither a parent nor a descendant of $i$.

\subsubsection{Estimating DAGs}  
A probability distribution may be represented by multiple DAGs that imply the same conditional independencies. The collection of all such DAGs is called a \textit{Markov equivalence class}. All DAGs in an equivalence class have the \textit{same skeleton} (undirected version of the DAG) and the \textit{same unshielded colliders} \citep{verma1990equivalence}, while they may differ in the orientation of other edges in the skeleton. An equivalence class can be represented by a completed partial DAG (CPDAG), a hybrid graph consisting of all directed edges common to all DAGs in the equivalence class and undirected edges wherever at least two DAGs have arrows in opposite directions.

Thus, in general, it is only possible to estimate the equivalence class rather than a unique DAG. A widely used approach to estimating the equivalence class is the PC algorithm \citep{spirtes1991algorithm}, which we outline in Algorithm~\ref{algo:pc}. The PC algorithm allows us to estimate the skeleton and the set of all unshielded colliders, which, together with the no-cycle constraint, is all we need to estimate the equivalence class of DAGs representing the multivariate distribution of interest. 
 
\begin{algorithm}[ht!]\small
{\bf Input}: Data $X^{(1)},...,X^{(n)}$; adjacency set $C=\{(i,j):i\neq j\}$; collection of empty separator sets $\{\mathcal{A}_{ij}\}_{i\neq j}$; significance level $\alpha$.
\begin{enumerate}
    \item {\sc Skeleton}\\
For $k=0,1,...,d$:\\
\hspace*{1mm} For every pair $(i,j)$ such that $\left|\mathcal{N}_{C}(i)\setminus\{j\}\right|\ge k$:\\ 
\hspace*{2mm} For every $A\subseteq \mathcal{N}_{C}(i)\setminus\{j\}$ such that $|A|=k$:\\ 
\hspace*{3mm} Test $H_0$: $ (X_i,X_j)|X_A$  indep. VS $H_1$: $(X_i,X_j)| X_A$ dep. \\ 
\hspace*{4mm} $H_0$ not rejected $\Rightarrow$ $C\equiv C\setminus\{(i,j),(j,i)\}$  \&  $\mathcal{A}_{ij}\equiv\mathcal{A}_{ij}\cup A$. 
 \item {\sc Unshielded Colliders}\\ 
 For every $(i,j)\notin C$:\\ 
\hspace*{1mm} For every $k\in\mathcal{N}_C(i)\cap \mathcal{N}_C(j)$:\\ 
\hspace*{2mm} If $k\notin\mathcal{A}_{ij}$, update $C\equiv C\setminus\{(k,i),(k,j)\}\cup\{(i,k),(j,k)\}$.
\item CPDAG\\ 
Assign arrows to undirected edges in $C$ without creating any new unshielded colliders or cycles.
\end{enumerate} 
{\bf Output}: CPDAG $=C$. 
\caption{PC algorithm}\label{algo:pc}
\end{algorithm} 

Algorithm~\ref{algo:pc} takes as input the data $X^{(1)},...,X^{(n)}$, a starting adjacency set $C=\{(i,j):i\neq j\}$, a starting collection of separator sets $\mathcal{A}=\{\mathcal{A}_{ij}\}_{i\neq j}$, where $\mathcal{A}_{ij}=\emptyset$, for all $i\neq j$, and a test size $\alpha$. In step 1, we find the skeleton by implementing a sequence of hypothesis tests on the independence of variable pairs $(X_i,X_j)$ conditionally on other variables ($X_A=(X_k)_{k\in A}$) that are neighbors with $i$ and/or $j$ in $C$. Each hypothesis test may be implemented in various ways. For example, if the data are multivariate Gaussian, as per Equation~\eqref{eq:fisherasymp}, we can use the test statistic $T_{ij,A}=|g(\hat\rho_{ij\mid A})|\sqrt{n-|A|-3}$, where $\hat\rho_{ij\mid A}$ is the sample partial correlation between $(X_i,X_j)$ given $X_A$ (i.e., $\hat\rho_{ij\mid A}$ is computed as in Equation~\eqref{eq:parcorhat} but with $\hat\Theta$ being the inverse of the sample covariance matrix $\hat\Sigma_{(i,j,A)\times (i,j,A)}$ relative to nodes $i,j,A$), and $g(\omega)$ is the Fisher transformation; we reject $H_0$ if $T_{ij}>z_{\alpha/2}$, where $z_{\alpha/2}=\Phi^{-1}(1-\alpha/2)$ is the $1-\alpha/2$ quantile of $N(0,1)$, and $\Phi^{-1}(\cdot)$ is the inverse of the c.d.f. of $N(0,1)$ in Equation~\eqref{eq:phiz}. We use this test in the data analysis of Section~\ref{sec:data}, where we apply Algorithm~\ref{algo:pc} to the normal scores (Equation~\eqref{eq:normalscore}) of the data. Implementing this test for all $(i,j)$ and all suitable potential separators $A$, the skeleton estimate is finally given by what remains in the set $C$. During this process, the collection of separator sets $\{\mathcal{A}_{ij}\}_{i\neq j}$ is also updated. 

In step 2, we identify all unshielded colliders. We look through all triples $(i,k,j)$ where $i,j$ are not connected in the skeleton, while $k$ is adjacent to both $i$ and $j$ and is not a separator of $(i,j)$ (i.e., $k\notin \mathcal{A}_{ij}$). Any such triple $(i,k,j)$ forms an unshielded collider at $k$. During this process, the adjacency set $C$ is updated to reflect the orientation (i.e., $i\rightarrow k \leftarrow j$) of the edges in the unshielded colliders. 

Finally, in step 3, we attempt to assign orientations to all remaining undirected edges in $C$ by simply applying the constraints that (i) no cycles and (ii) no additional unshielded colliders may be created. The output of the Algorithm is the CPDAG $C$, which may still contain undirected edges for all those pairs $(i,j)$ where a unique orientation could not be determined.

\section{Analysis of exoplanets and host stars data}\label{sec:data}
In this section, we apply graphical models to estimate the dependence structure among multiple astrophysical variables characterizing exoplanets and their host stars. All computations were implemented with the R package \texttt{astroggm}, which we developed and made available on GitHub.    

The exoplanet and host star data we analyze were taken from the NASA Exoplanet Archive and consist of astrophysical measurements from confirmed exoplanets identified through a variety of detection methods. We focus our analysis on the $d=10$ quantitative variables listed in Table~\ref{table:exoplanets}. To apply the methods presented in Section~\ref{sec:gms}, all data vectors must be complete. Thus, all cases containing any missing values were removed, yielding a total number of $n=1067$ exoplanets. Moreover, we apply the nonparametric transformation in Equation~\eqref{eq:normalscore} to all variables before pursuing any analysis.

\begin{table}[ht!]
\centering\footnotesize
{ \begin{tabular}{p{1.3cm}|p{6.9cm}}
\hline
{\bf P.mass} & Planet mass relative to Earth.\\
{\bf P.orb.per} & Planet orbital period.
\\
{\bf P.orb.ecc} & Planet orbital eccentricity.
\\
{\bf S.temp} & Star surface temperature. \\
{\bf S.radius} & Star radius relative to the Sun. \\
{\bf S.mass} & Star mass relative to the Sun. \\
{\bf S.metal} & Star abundance of elements heavier than \textit{H} and \textit{He}. \\
{\bf S.gravity} & Star surface gravitational pull. \\
{\bf S.dist} & Star distance from Earth to the star.
\\
{\bf S.gaia} & Star apparent magnitude in the Gaia G-band. \\
\hline
\end{tabular}
}
\caption{Variables in the exoplanet and host star data.}\label{table:exoplanets}
\end{table}

\begin{figure*}[ht!]
    \centering
\includegraphics[width=.479\linewidth]{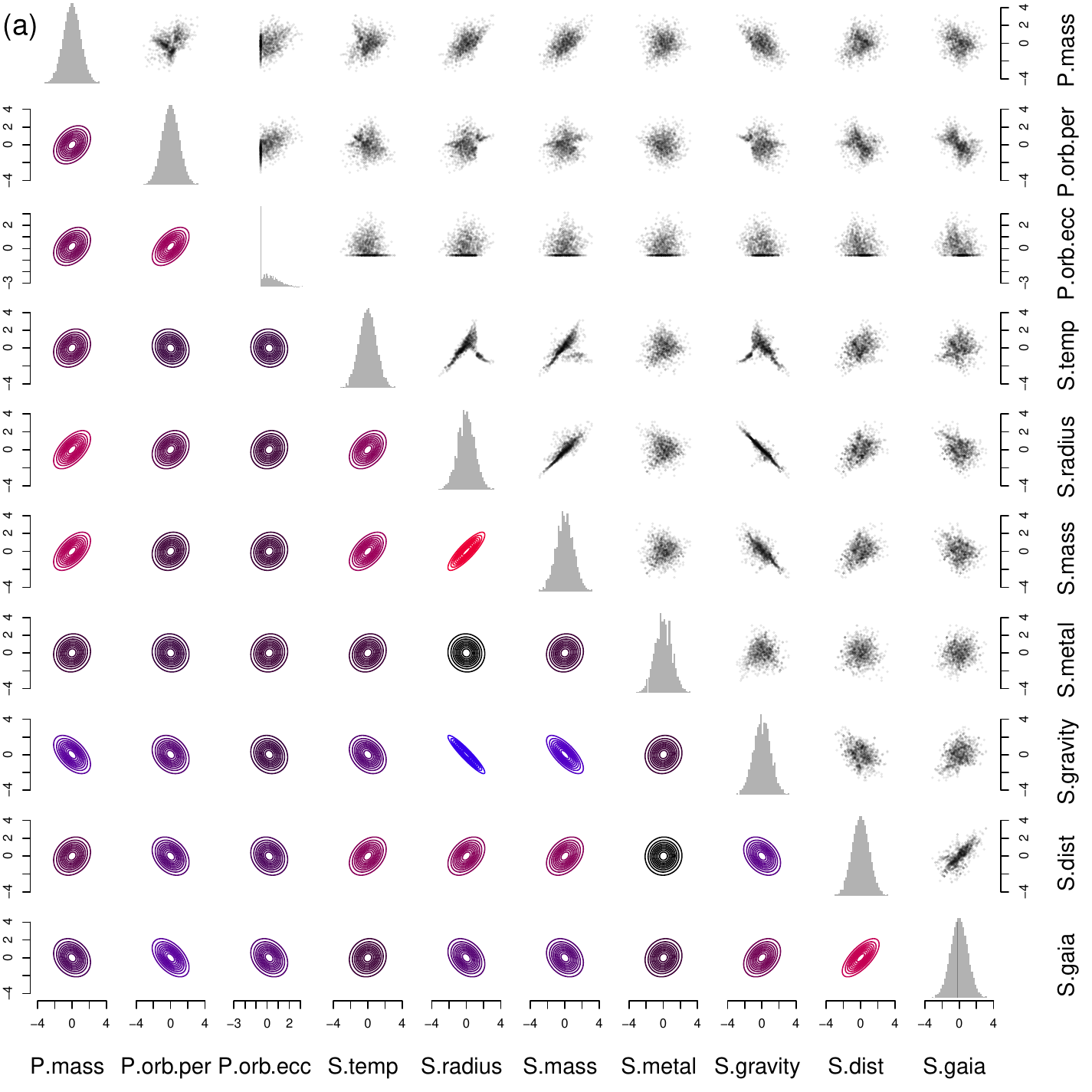}\hfill \includegraphics[width=.48\linewidth]{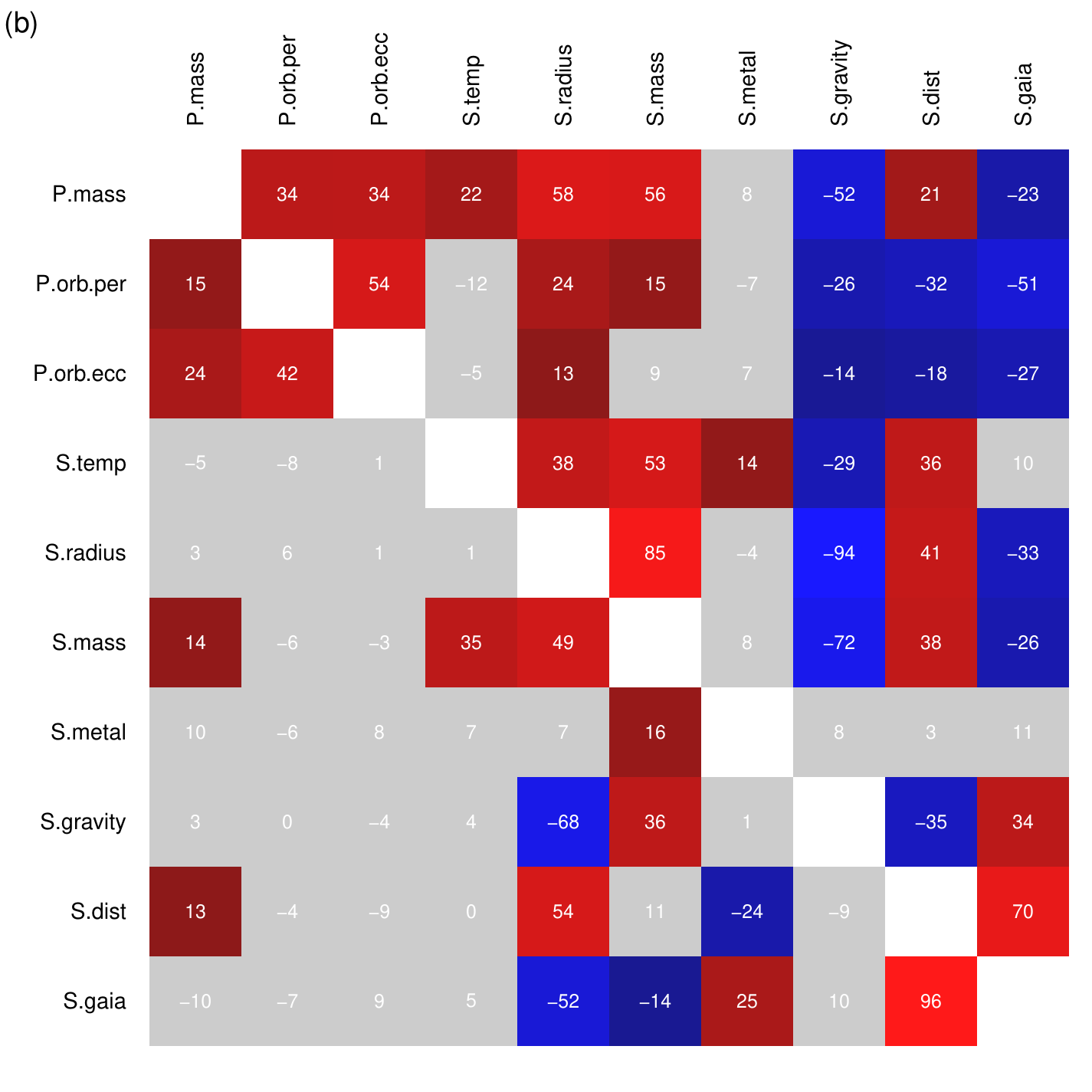}
    \caption{(a): Histograms of the ten variables listed in Table~\ref{table:exoplanets} (diagonal plots) standardized via Equation~\eqref{eq:normalscore}; scatterplots of each variable versus every other (upper off-diagonals); and contour plots of estimated bivariate Gaussian distributions (lower diagonals). (b): Heatmap of the sample marginal correlations $\hat r_{ij}$ (upper diagonals; Equation~\eqref{eq:corrmle}) and sample partial correlations $\hat\rho_{ij}$ (lower diagonals; Equation~\eqref{eq:parcorhat}) of the ten variables, where blue colors denote negative correlations, red colors denote positive correlations, and grey color denotes correlations not significantly different from zero (FWER$\le$1\%, Bonferroni correction). Values of the correlations and partial correlations are shown as rounded signed percentages.
    }
\label{fig:exoplanetsEDA}
\end{figure*}

\begin{table*}[t!]
\centering\scriptsize \addtolength{\tabcolsep}{-0.19em}
\begin{tabular}{r|cccccccccc}
 & \textbf{P.mass} & \textbf{P.orb.per} & \textbf{P.orb.ecc} & \textbf{S.temp} & \textbf{S.radius} & \textbf{S.mass} & \textbf{S.metal} & \textbf{S.gravity} & \textbf{S.dist} & \textbf{S.gaia} \\ 
  \hline
\textbf{P.mass} & $\star$ &  (24,\textbf{34},44) &  (24,\textbf{34},44) &  (11,\textbf{22},32) &  (49,\textbf{58},65) &  (47,\textbf{56},63) &  (-4,{8},19) &  (-60,\textbf{-52},-43) &  (10,\textbf{21},31) &  (-33,\textbf{-23},-12) \\ 
  \textbf{P.orb.per} &  (3,\textbf{15},26) & $\star$ &  (45,\textbf{54},61) &  (-23,{-12},0) &  (13,\textbf{24},34) &  (4,\textbf{15},26) &  (-18,{-7},5) &  (-36,\textbf{-26},-15) &  (-42,\textbf{-32},-22) &  (-59,\textbf{-51},-43) \\ 
  \textbf{P.orb.ecc} &  (13,\textbf{24},35) &  (32,\textbf{42},51) & $\star$ &  (-16,{-5},6) &  (2,\textbf{13},24) &  (-3,{9},20) &  (-4,{7},19) &  (-25,\textbf{-14},-3) &  (-29,\textbf{-18},-7) &  (-37,\textbf{-27},-16) \\ 
  \textbf{S.temp} &  (-16,{-5},7) &  (-19,{-8},3) &  (-10,{1},12) & $\star$ &  (28,\textbf{38},48) &  (44,\textbf{53},60) &  (3,\textbf{14},25) &  (-39,\textbf{-29},-18) &  (26,\textbf{36},46) &  (-2,{10},21) \\ 
  \textbf{S.radius} &  (-9,{3},14) &  (-6,{6},17) &  (-10,{1},13) &  (-11,{1},12) & $\star$ &  (82,\textbf{85},88) &  (-15,{-4},7) &  (-95,\textbf{-94},-92) &  (31,\textbf{41},50) &  (-42,\textbf{-33},-22) \\ 
  \textbf{S.mass} &  (3,\textbf{14},25) &  (-17,{-6},5) &  (-14,{-3},8) &  (24,\textbf{35},44) &  (40,\textbf{49},57) & $\star$ &  (-3,{8},19) &  (-77,\textbf{-72},-66) &  (28,\textbf{38},47) &  (-36,\textbf{-26},-15) \\ 
  \textbf{S.metal} &  (-1,{10},21) &  (-17,{-6},5) &  (-3,{8},20) &  (-5,{7},18) &  (-4,{7},18) &  (5,\textbf{16},27) & $\star$ &  (-3,{8},19) &  (-9,{3},14) &  (0,{11},22) \\ 
  \textbf{S.gravity} &  (-8,{3},14) &  (-11,{0},12) &  (-16,{-4},7) &  (-7,{4},15) &  (-74,\textbf{-68},-62) &  (26,\textbf{36},46) &  (-11,{1},12) & $\star$ &  (-45,\textbf{-35},-25) &  (24,\textbf{34},43) \\ 
  \textbf{S.dist} &  (2,\textbf{13},24) &  (-15,{-4},8) &  (-21,{-9},2) &  (-11,{0},11) &  (45,\textbf{54},61) &  (-1,{11},22) &  (-34,\textbf{-24},-13) &  (-20,{-9},2) & $\star$ &  (64,\textbf{70},76) \\ 
  \textbf{S.gaia} &  (-21,{-10},1) &  (-19,{-7},4) &  (-3,{9},20) &  (-6,{5},16) &  (-60,\textbf{-52},-43) &  (-24,\textbf{-14},-2) &  (14,\textbf{25},35) &  (-1,{10},21) &  (95,\textbf{96},97) & $\star$
\end{tabular}
\caption{Sample marginal correlations $\hat r_{ij}$ (upper diagonals; Equation~\eqref{eq:corrmle}) and sample partial correlations $\hat\rho_{ij}$ (lower diagonals; Equation~\eqref{eq:parcorhat}) as rounded signed percentages with 99$\%$ Bonferroni corrected confidence intervals  (Equation~\eqref{eq:fisherconfint},  $\alpha=0.01/\binom{10}{2}$). Values that are significantly different from zero are in bold face.}\label{table:exoplanetsempirical} 
\end{table*}

\begin{figure*}[ht!]
\includegraphics[width=1\textwidth]{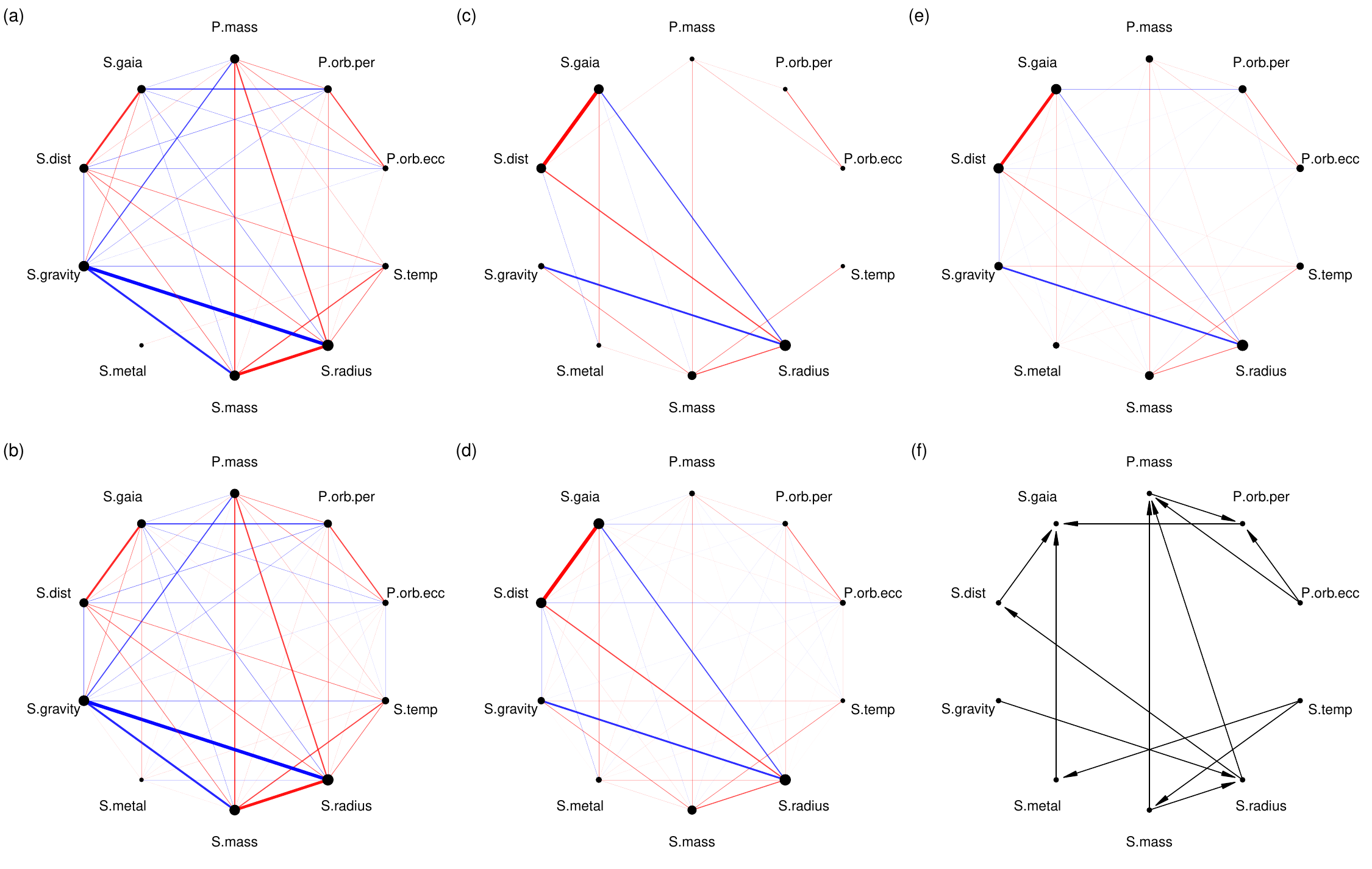}
    \caption{
    (a): Marginal correlation graph based on 1\% FWER control via Bonferroni correction (Table~\ref{table:exoplanetsempirical}, upper diagonals). (b): Marginal correlation graph based on 1\% FDR control. (c): Partial correlation (conditional dependence) graphs based on 1\% FWER control via Bonferroni correction (Equation~\eqref{eq:bonferroniedgeset}; Table~\ref{table:exoplanetsempirical}, lower diagonals). (d): Partial correlation graph based on 1\% FDR control  (Equation~\eqref{eq:fdredgeset}). 
    (e): Partial correlation graph based on GLASSO (Equation~\eqref{eq:glassoE}) with EBIC tuning parameter selection (Equation~\eqref{eq:ebic}). In all graphs in (a)--(e), blue edges denote negative correlations, red edges denote positive correlations, the thickness of each edge is proportional to the magnitude of the correlation, and the radius of each node is proportional to the sum of the magnitudes of the correlations relative to the node. (f): CPDAG obtained via PC algorithm (Algorithm~\ref{algo:pc}, $\alpha=0.01$). 
    }
\label{fig:exoplanetsgraphs}
\end{figure*}

In Figure~\ref{fig:exoplanetsEDA}(a), we show the histograms of the ten variables standardized via Equation~\eqref{eq:normalscore} (diagonal plots), scatterplots of each variable versus every other (upper off-diagonals), and contour plots of estimated bivariate Gaussian distributions. In Figure~\ref{fig:exoplanetsEDA}(b), we show the heatmap of the sample marginal correlations (upper diagonals; Equation~\eqref{eq:corrmle}) and sample partial correlations (lower diagonals; Equation~\eqref{eq:parcorhat}) of the ten variables, where blue colors denote negative correlations, red colors denote positive correlations, and the gray color denotes correlations that are not significantly different from zero (FWER$\le$1\%, Bonferroni correction), as per Table~\ref{table:exoplanetsempirical}. In this table, we report the sample marginal correlations (upper diagonals) and the sample partial correlations (lower diagonals) as percentages, with 99$\%$ Bonferroni corrected confidence intervals (Equation~\eqref{eq:fisherconfint}, with $\alpha=0.01/\binom{10}{2}$). 

In Figures~\ref{fig:exoplanetsgraphs}(a)--(b), we show the estimated marginal correlation (marginal dependence) graphs based on 1\% FWER control via Bonferroni correction (Table~\ref{table:exoplanetsempirical}, upper diagonals) and 1\% FDR control, respectively.  
In Figures~\ref{fig:exoplanetsgraphs}(c)--(e), we show the partial correlation (conditional dependence) graphs based on 1\% FWER control via Bonferroni correction (Equation~\eqref{eq:bonferroniedgeset}; Table~\ref{table:exoplanetsempirical}, lower diagonals), 1\% FDR control (Equation~\eqref{eq:fdredgeset}), and GLASSO (Equation~\eqref{eq:glasso}) selected via EBIC (Equation~\eqref{eq:ebic}), respectively. In all these graphs, blue edges denote negative correlations, red edges denote positive correlations, the thickness of each edge is proportional to the magnitude of the correlation, and the radius of each node is proportional to the sum of the magnitudes of the correlations relative to the node. Finally, in Figure~\ref{fig:exoplanetsgraphs}(f), we show the CPDAG representing the Markov equivalence class estimated via the PC algorithm (Algorithm~\ref{algo:pc}, $\alpha=0.01$). Since no undirected edge is present, the estimated Markov equivalence class contains only one DAG.

A considerable number of dependencies can be noticed, both positive and negative, spanning stellar, planetary, and orbital parameters. 
The partial correlation graphs in Figures~\ref{fig:exoplanetsgraphs}(c)--(e) are sparser than the marginal correlation graphs in Figures~\ref{fig:exoplanetsgraphs}(a)--(b), indicating that numerous apparent pairwise relationships weaken or disappear once the effects of other variables are accounted for in the graphical model. Indeed, conditional dependencies provide a more complete and reliable explanation of the dependence structure among variables. The graph in Figure~\ref{fig:exoplanetsgraphs}(c) is the sparsest (16 edges) of the three partial correlation graphs, as  Bonferroni multiple hypothesis testing correction is known to be very conservative. 

Most of the identified conditional dependencies in Figures~\ref{fig:exoplanetsgraphs}(c)--(e), especially the strongest ones that are present in all graphs, correctly reproduce several established astrophysical principles. 
The negative partial correlation between stellar surface gravity (S.gravity) and stellar radius (S.radius), and the positive partial correlation between stellar surface gravity and stellar mass (S.mass) are consistent with the \textit{surface gravity formula} S.gravity $\propto$ S.mass/S.radius$^2$ \citep{carroll2017introduction}.  
The positive partial correlations connecting stellar mass (S.mass) to stellar radius (S.radius) and stellar effective temperature (S.temp) correspond to well-established \textit{main-sequence scaling} relations: more massive stars are larger (S.radius $\approx$ S.mass$^a$, $0.6\le a \le 0.8$) and hotter (S.temp $\approx$ S.mass$^{0.5}$) \citep{kippenhahn1990stellar}. 

The positive partial correlations connecting planet mass (P.mass) to stellar mass (S.mass) and stellar metallicity (S.metal) align with the \textit{core accretion model} \citep{laughlin2004core}: more massive stars with higher metallicity possess heavier protoplanetary disks, providing a larger accumulation of solids to build massive planets. Our results are also consistent with \cite{lozovsky2021more}, who found that more massive stars host planets that are systematically larger, and with \cite{owen2018metallicity}, who found that metal-rich stars favor giant planet formation. 

The positive partial correlation between apparent magnitude (S.gaia) and stellar distance (S.distance) reflects the \textit{distance–modulus relation} \citep{carroll2017introduction}, which explains why more distant stars appear fainter. Finally, the positive partial correlation between apparent magnitude (S.gaia) and stellar metallicity (S.metal) is consistent with \textit{stellar opacity theory} \citep{kippenhahn2012stellar}, whereby a star's chemical metals alter its atmospheric absorption, thereby shifting its perceived brightness in the Gaia photometric filters \citep{prusti2016gaia}. 

Also the dependence structure represented by the CPDAG in Figure~\ref{fig:exoplanetsgraphs}(f) is mostly consistent with the aforementioned astrophysical principles. 
The link S.mass~$\to$~P.mass aligns with the core accretion model \citep{laughlin2004core} and the results of  \cite{lozovsky2021more}. The unshielded collider  S.mass~$\rightarrow$~S.radius~$\leftarrow$~S.gravity is consistent with the law of gravitational acceleration at a surface \citep{carroll2017introduction}. The link S.temp~$\rightarrow$~S.mass is consistent with the main sequence relation S.temp $\approx$ S.mass$^{0.5}$ \citep{kippenhahn1990stellar}. Apparent magnitude (S.gaia) is a descendant of all other nodes. In particular, the unshielded collider S.metal~$\rightarrow$~S.gaia~$\leftarrow$~S.distance reflects the distance–modulus relation \citep{binney1998galactic} and  stellar opacity theory \citep{kippenhahn2012stellar}. Finally, the full connectivity between planet mass (P.mass), planet orbital period (P.orb.per), and planet orbit eccentricity (P.orb.ecc) is reasonable, with planet mass being the one directly affected by stellar parameters.

\section{Discussion}\label{sec:disc}
In this article, we explored the application of probabilistic graphical models in astronomy. Graphical models allow us to efficiently study the complex multivariate relationships among numerous variables. In an undirected graph, two nodes are disconnected if and only if the two variables they represent are independent, conditionally on all other variables. In a directed acyclic graph, any variable, conditionally on its parents, depends only on its parents and descendants. 

After providing an overview of undirected and directed probabilistic graphical models, we applied these frameworks to detect dependencies among variables concerning exoplanets and host stars. Most of the detected dependencies reflected established planetary and stellar astrophysical principles, such as main-sequence scaling relations \citep{kippenhahn1990stellar}, the surface gravity formula \citep{carroll2017introduction}, the core accretion model \citep{laughlin2004core}, the distance–modulus relation \citep{carroll2017introduction}, and stellar opacity theory \citep{kippenhahn2012stellar}. Conditional dependence graphs capture the dependence structure of variables more precisely than marginal correlations, which can often be misleading. For example, in our data analysis, stellar surface gravity (S.gravity) and stellar mass (S.mass) exhibited a strong negative marginal correlation and a positive partial correlation, with the latter being the one consistent with astrophysical theory.

Our results suggest that graphical models not only can accurately retrieve known relations but also may serve as powerful exploratory tools for gaining insight into astronomical variables that are not yet well understood. Therefore, while still not widely applied in astronomy, we expect probabilistic graphical models to become a common statistical tool in future astronomical research.

While in this paper we presented various well-established approaches to graphical model estimation, numerous alternative methods exist in the literature. These include extensions and Bayesian variants of GLASSO  \citep{wang2012bayesian,chandrasekaran2012,vinci2018adjusted,vinci2018adjustedB}, other forms of regularization  \citep{meinshausen2006high,cai2011constrained}, and methods for dependence estimation from structurally incomplete data  \citep{vinci2019graph,vinci2024linkedfactoranalysis,steneman2024covariance}. The application of these methods in astronomy will be the focus of our future research.

\section*{Declaration of competing interest}
The authors declare that they have no known competing financial interests or personal relationships that could have appeared to influence the work reported in this paper.

\section*{Data and code availability}
The R package \texttt{astroggm} used to implement the analyses is available on GitHub at \href{https://github.com/gvincistat}{ https://github.com/gvincistat}. The data set used in our analyses can also be found at \href{https://github.com/gvincistat}{ https://github.com/gvincistat}, and was originally retrieved from the NASA Exoplanet Archive \href{https://exoplanetarchive.ipac.caltech.edu/}{https://exoplanetarchive.ipac.caltech.edu/}.

\section*{Acknowledgements}
This research was funded by the Beutter Family Endowment for Excellence in Applied Mathematics and Computer Science and the Sheila Flynn Endowment for Excellence at the University of Notre Dame.

\bibliographystyle{apalike} 
\bibliography{bibliography.bib}

\end{document}